\documentclass[prl,twocolumn,showpacs,superscriptaddress]{revtex4}
\usepackage{graphicx,amssymb,amsmath}
\begin{document}

\title{Non-linear electric transport in graphene: quantum quench dynamics and the Schwinger mechanism}
\author{Bal\'azs D\'ora}
\email{dora@pks.mpg.de}
\affiliation{Max-Planck-Institut f\"ur Physik komplexer Systeme, N\"othnitzer Str. 38, 01187 Dresden, Germany}
\author{Roderich Moessner}
\affiliation{Max-Planck-Institut f\"ur Physik komplexer Systeme, N\"othnitzer Str. 38, 01187 Dresden, Germany}
\date{\today}

\begin{abstract}
We present a unified view of electric transport in undoped clean graphene for finite electric field.  
The weak field results agree with the Kubo approach. For 
strong electric field, the current increases non-linearly with the electric field as $E^{3/2}$. 
As the Dirac point is moved around in reciprocal space by the field, 
excited states are generated, in a way analogous to the generation of defects in a quench  through a quantum 
critical point. 
These results are also analyzed in terms of Schwinger's pair production and Landau-Zener tunneling.
An experiment for cold atoms in optical lattices is proposed to test these ideas.
\end{abstract}


\maketitle


The discovery of graphene, a single sheet of carbon atoms in a honeycomb lattice (HCL) 
has triggered intense research recently\cite{novoselov1,castro07} not only because 
of its potential application in future electronic devices, but also because of its fundamental physical properties: its 
quasiparticles are governed by the two-dimensional Dirac equation, and 
exhibit a variety of compelling (pseudo)relativistic phenomena such as the
 unconventional quantum Hall 
effect\cite{novoselov2}, a (possibly universal) minimal conductivity at vanishing carrier concentration\cite{ziegleropt}, 
Klein tunneling in p-n junctions\cite{cheianovpn,beenakker} and Zitterbewegung\cite{katsnelson3}.

Quantum transport and non-linear responses driven by finite
external fields represent a genuine non-equilibrium phenomenon, 
giving rise to e.g. dielectric breakdown or Bloch oscillations\cite{dahan}.
The quantum aspect of these effects is particularly pronounced in reduced dimensions.
Therefore, two-dimensional Dirac electrons in finite electric fields, the subject of this work, provide a fascinating setting for 
studying these issues.

A simple  picture of electronic transport in a finite electric field is drift transport as considered
 by Drude: carriers move ballistically ($p=eEt$) until
they change their momentum by a scattering process, replacing the time $t$  by the
appropriate scattering time.
The special features of Dirac electrons relevant for transport in finite field include:
(i) their velocity is pinned to the "light cone" Fermi velocity, $v_F$,
(ii) relativistic particles undergo pair production in strong electric fields, 
as predicted by 
Schwinger\cite{schwinger}, and 
(iii) a uniform electric field 
modifies locally the geometry of the Fermi surface by moving the 
Dirac point around in momentum space (Eq. \eqref{eigenenergy}).
Since massless Dirac electrons can be thought of as being
critical, this can lead to the production of excited states, and should leave its fingerprints on transport in finite electric fields.

The Landau-Zener (LZ) dynamics, describing the (avoided) level crossing in a two level system\cite{vitanov}, represents the natural language to discuss Klein tunneling\cite{beenakker,allor} in 
graphene, 
and is ultimately connected to defect formation and quench dynamics through quantum critical 
points\cite{damski}, described by the Kibble-Zurek mechanism\cite{kibble,zurek} of non-equilibrium phase transitions. 
Applying these ideas to graphene allows us to analyze the real time dynamics of the current, after switching on the electric field, and to identify the various
crossovers (summarized in Table \ref{tab1}).
Electric transport depends sensitively on the frequency, temperature, electric field and scattering 
rate ($\omega$, $T$, $E$, $\Gamma$), and the obtained current depends strongly on how the 
$(\omega,T,E,\Gamma)\rightarrow 0$ limit is 
taken\cite{green,ziegleropt}. Our results follow from taking the $\omega=T=0$ limits in a finite electric field.

\begin{table}[h!]
\centering
\begin{tabular}{|c|c|c|}
\hline
classical & Kubo & Schwinger/Kibble-Zurek \\
 $t\ll h/W$ & $h/W\ll t \ll \sqrt{\hbar/v_FeE}$ & $\sqrt{\hbar/v_FeE} \ll t\ll t_{Bloch}$\\
\hline
$j_x\sim Et$  & $j_x\sim E$ & $j_x\sim tE^{3/2}$ \\
\hline
\end{tabular}
\caption{Temporal evolution of the non-equilibrium current for clean graphene. Bloch oscillations show up for
$t\gtrsim t_{Bloch}\sim\hbar/eaE$\cite{dahan} with $a$ the HCL constant.
\label{tab1}}
\end{table}

We focus on the 2+1 dimensional Dirac equation in a uniform, constant electric field in the $x$ direction, switched on at $t=0$, 
through 
a time dependent vector potential as
${\bf A}(t)=(A(t),0,0)$ with $A(t)=Et\Theta(t)$.
The resulting time dependent Dirac equation, describing low energy excitations around the $K$ point in the Brillouin zone for clean 
graphene, 
is written as 
\begin{gather}
\begin{split}
H=v_F[\sigma_x(p_x-eA(t))+\sigma_yp_y],\\
i\hbar\partial_t\Psi_p(t)=H\Psi_p(t),
\end{split}
\label{diraceq}
\end{gather}
where $v_F\simeq 10^6$~m/s is the Fermi velocity of grap\-he\-ne, and the Pauli matrices ($\sigma$) arise from the two 
sublattices\cite{castro07} of the HCL. 
Due to this (pseudo)spin structure, Eq. \eqref{diraceq} represents a natural platform to study 
LZ dynamics as well.
It is convenient to perform a time dependent unitary transformation first\cite{cohen}, which diagonalizes $H$, and brings us to the 
adiabatic basis in the LZ language\cite{vitanov} as
\begin{gather}
U=\frac{1}{\sqrt{2}}\left(
\begin{array}{cc}
\exp(-i\varphi/2)  & \exp(-i\varphi/2)\\
\exp(i\varphi/2) & -\exp(i\varphi/2)
\end{array}\right),
\end{gather}
$U^+HU=\sigma_z\epsilon_p(t)$, where 
\begin{gather}
\epsilon_p(t)=v_F\sqrt{(p_x-eA(t))^2+p_y^2},\label{eigenenergy}
\end{gather}
and $\tan \varphi=p_y/(p_x-eA(t))$.
With this, the transformed time dependent Dirac equation is given by
\begin{gather}
i\hbar\partial_t\Phi_p(t)=\left[\sigma_z\epsilon_p(t)-\sigma_x\frac{\hbar 
v_F^2p_yeE}{2\epsilon^2_p(t)}\right]\Phi_p(t),
\label{diractrans}\\
\Phi_p^T(t=0)=(0,1),
\end{gather}
where $\Psi_p(t)=U\Phi_p(t)$, and the off diagonal terms in the Hamiltonian arise due to the explicit time dependence of the unitary 
transformation ($-iU^+\partial_tU$), and the initial condition corresponds to zero temperature and half filling. The main advantage of the unitary 
transformation is that the resulting equation clearly 
distinguishes between positive and negative energy states. 

The current operator in the original basis is obtained through the equation of motion as $j_x=-ev_F\sigma_x$. After the 
unitary transformation, it reads as
$j_x=-ev_F(\sigma_z\cos\varphi+\sigma_y\sin\varphi)$.
In the presence of the electric field, the expectation value of the current is finite. By denoting 
$\Phi_p^T(t)=(\alpha_{p}(t),\beta_{p}(t))$, 
\begin{gather}
\langle 
j_x\rangle_p(t)=-ev_F\left[\cos\varphi(|\alpha_{p}(t)|^2-|\beta_{p}(t)|^2)+\right.\nonumber\\
\left.+2\sin\varphi\textmd{Re}(i\alpha_{p}(t)\beta^*_{p}(t))\right].
\end{gather}
The first term is the current from particles residing on the upper or lower Dirac cone, while the second one describes interference 
between them, and is responsible for Zitterbewegung. Using QED terminology, the first and second term is referred to as conduction and 
polarization current, respectively\cite{tanji}. In condensed matter, these are called intraband and interband 
contributions, respectively.
Due to charge conservation, $|\alpha_{p}(t)|^2-|\beta_{p}(t)|^2=2|\alpha_{p}(t)|^2-1$.
The interference correction also simplifies since 
\begin{gather}
\partial_t|\alpha_{p}(t)|^2=2\textmd{Re}(\alpha_{p}(t)\partial_t\alpha^*_{p}(t)).
\end{gather}
By using the transformed Hamiltonian, Eq. \eqref{diractrans}, 
\begin{gather}
\hbar\partial_t\alpha^*_{p}(t)=i\epsilon_p(t)\alpha^*_{p}(t)-i\frac{\hbar v_F^2p_yeE}{2\epsilon^2_p(t)}\beta^*_p(t),
\end{gather}
consequently
\begin{gather}
\partial_t|\alpha_{p}(t)|^2=-\frac{v_F^2p_yeE}{\epsilon^2_p(t)}\textmd{Re}(i\alpha_p(t)\beta^*_p(t)),
\end{gather}
since Re$(i|\alpha_p(t)|^2)=0$.
As a result, the expectation value of the current only requires the knowledge of $n_p(t)=|\alpha_p(t)|^2$ as
\begin{gather}
\langle
j_x\rangle_p(t)=-ev_F\left[\frac{v_F(p_x-eEt)}{\epsilon_p(t)}(2n_{p}(t)-1)-\right.\nonumber\\
\left.-2\frac{\epsilon_p(t)}{v_FeE}\partial_t 
n_p(t)\right].
\label{current}
\end{gather}
The term independent of $n_p(t)$, namely $ev_F^2(p_x-eEt)/\epsilon_p(t)$ vanishes at half filling after momentum integration. In QED, 
this originates from 
charge conjugation symmetry\cite{tanji}, while in graphene, it is obtained by taking the full honeycomb lattice into 
account as in  Ref. \onlinecite{lewkowicz}.

For $t<0$, the upper/lower Dirac cone is empty/fully occupied. The quantity $n_p(t)$ measures the number 
of particles created by the electric field in the upper cone through Schwinger's pair 
production\cite{schwinger}.
In graphene, instead of particle-antiparticle pairs, electron-hole pairs are created.
Therefore, the basic quantity to determine transport through graphene is $n_p(t)$.
We start by analyzing its behaviour at weak electric fields perturbatively. In this case, we can set $E=0$ in Eq. \eqref{diractrans} 
except in the 
numerator of the off-diagonal terms, and obtain
\begin{equation}
n_p(t)=\frac{(eE\hbar p_y)^2}{4v_F^2|p|^6}\sin^2\left(\frac{v_F|p|t}{\hbar}\right),
\label{nppert}
\end{equation}
which is valid except in the close vicinity of the Dirac point (i.e. $|p|\gg eEt$), and $|p|=\sqrt{p_x^2+p_y^2}$.
Plugging this into Eq. \eqref{current}, the first term is already second order in the electric field, and does not contribute to 
linear response.
The second (polarization) term gives, taking valley and spin degeneracies into account
\begin{equation}
\langle j_x\rangle=\frac{e^2E}{2\pi\hbar}\int\limits_0^\infty\textmd{d}p\frac{\sin(2v_Fpt/\hbar)}{p}=\frac{e^2}{4\hbar}E,
\end{equation}
and the dc conductivity is $\sigma=j/E=e^2\pi/2h$, in accordance with Ref. \cite{lewkowicz}.
This is the value of the ac conductivity at finite frequencies obtained from the Kubo formula\cite{ziegleropt,castro07} and measured 
also\cite{opticalgeim}, and since the model does not contain any additional 
energy scale, 
which would change the value of the ac response down to $\omega\rightarrow 0$, the same value for the dc conductivity sounds 
plausible.
 Within our approach, the small field response is 
dominated by Zitterbewegung corrections.
The ultrashort time transient response ($tW\ll h$ with $W$ the bandwidth) is fully classical.
Expanding Eq. \eqref{nppert},
 we obtain
\begin{equation}
\langle j_x\rangle_p(t)=e^2v_F\frac{p_y^2}{|p|^3}Et,
\label{currentclassical}
\end{equation}
independent of $\hbar$. The current rises linearly with time after the switch on as $\langle j_x\rangle(t)=4e^2EWt/h^2$.
The very same result follows from a classical Hamiltonian, $H_{cl}=v_F\sqrt{(p_x-eEt)^2+p_y^2}$.
The Hamilton equation is
\begin{gather}
\partial_t x=\frac{\partial H_{cl}}{\partial p_x}=\frac{v_F^2(p_x-eEt)}{H_{cl}},
\label{diracvelocity}
\end{gather}
which gives for the classical current, $j_{cl}(p,t)=-e\partial_t x$, at short times as in Eq. \eqref{currentclassical}.
Dirac particles can therefore be accelerated as $\partial^2_tx=eE/m_{xx}$ at short times, in accord with Newton's 
equation after defining their effective mass as $1/m_{xx}=\partial^2H/\partial p_x^2=v_Fp_y^2/|p^3|$. 

\begin{figure}[h!]

{\includegraphics[width=3.6cm,height=3.8cm]{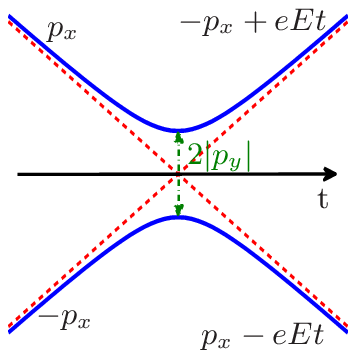}}
\hspace*{2mm}
{\includegraphics[width=4cm,height=4cm]{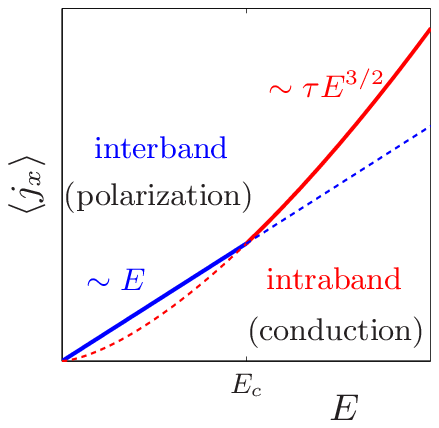}}

\caption{(Color online) Left panel: visualization of the temporal evolution of the LZ dynamics. Right panel: schematic 
picture of the current-electric field characteristics for graphene. Interband transitions are 
overwhelmed by intraband ones with increasing electric field, and the character of the measured current changes from polarization to 
conduction contribution.
\label{nlfig}}
\end{figure}

For the general time and electric field dependence, Eq. \eqref{diraceq} can be solved analytically\cite{gavrilov,tanji} using 
the parabolic cylinder functions, which do not immediately yield a transparent analytical expression  for the electric 
current for arbitrary electric field and time. To investigate the strong field, 
long time (specified in Eq. \eqref{crosstime}) response of Dirac electrons, we 
 use the asymptotic expansion of these eigenfunctions\cite{tanji,cohen}, or equivalently we can rely on the WKB 
approach\cite{casher} to 
determine $n_p(t)$ through the barrier penetration factor, similarly to 
narrow gap semiconductors\cite{aronov}.
As a result, we get 
\begin{equation}
n_p(t)=\Theta(p_x)\Theta(eEt-p_x)\exp\left(-\frac{\pi v_Fp_y^2}{\hbar eE}\right),
\label{pairprod}
\end{equation}
which is the celebrated pair production rate by Schwinger\cite{schwinger,tanji}, a manifestation of 
Klein tunneling\cite{beenakker}, and also the LZ transition 
probability\cite{vitanov} between the initial and 
final levels. More precisely, the conditions for applicability are $(p_x,eEt-p_x) \gg |p_y|$. 
This expression can be transparently understood invoking LZ physics. Two levels at $\pm p_x$, weakly coupled by $p_y$ 
level cross with time, ending up at $\pm (p_x-eEt)$. The transition is completed when both the initial and final levels are
well separated, in which case the mixing between them is given by Eq. \eqref{pairprod}, as plotted in Fig. \ref{nlfig}.

Putting Eq. \eqref{pairprod} into Eq. \eqref{current}, the current is dominated by the conduction (intraband 
part) as
\begin{equation}
\langle j_x\rangle(t)=\frac{2e^2E}{\pi^2\hbar}\sqrt{\frac{v_FeEt^2}{\hbar}},
\label{schwingercurrent}
\end{equation}
which increases linearly with time, similarly to normal electrons in a parabolic band. However, the origin of the time 
dependence is completely 
different: it stems from the increasing number of  pairs due to pair production \`a la Schwinger, each contributing with the same 
velocity $v_F$, as opposed to the continuously accelerated fixed number of normal electrons in strong fields.

The total number of particles and holes created is
\begin{equation}
N(t)=\frac{2}{\hbar^2\pi^2}\int \textmd{d}{\bf p}n_p(t)=\frac{2eE}{\pi^2 v_F \hbar}\sqrt{\frac{v_FeEt^2}{\hbar}},
\label{defectdensity}
\end{equation}
which leads to Eq. \eqref{schwingercurrent} via $\langle j_x\rangle(t)=ev_FN(t)$, and is related to the 
quench dynamics through a quantum critical point (QCP)\cite{dziarmaga} as follows:
Eq. \eqref{diraceq} can be diagonalized at every instant  with eigenenergies in Eq. \eqref{eigenenergy}:
 the Dirac point moves continuously in momentum space with location ${\bf p}=(eEt,0)$, which results 
in defect (excitation) production.
The spectra from Eq. \eqref{eigenenergy} can be considered as an ensemble of 1+1 dimensional initially 
gapped systems (labeled by $p_x$) driven through a QCP. The initial energy 
gap is given by $v_F|p_x|$, the one dimensional momentum is $p_y$, and the 
quench is applied as $v_F(p_x-eEt)$. For a given $p_x$, during the temporal evolution, the gap vanishes 
at the instant $t=p_x/eE$, which defines the QCP, and reappears with increasing time. The dynamics close 
to the QCP (characterized by $d=z=\nu=1$\cite{dziarmaga}) is necessarily non-adiabatic (impulse) due to the divergence 
of the relaxation time and the finite quench time $\sim 1/eE$\cite{damski}.
The Kibble-Zurek  mechanism\cite{kibble,zurek} 
predicts a scaling form for the defect formation as\cite{polkovnikov} 
$E^{d\nu/(z\nu+1)}=E^{1/2}$ for a given 1+1 dimensional system.

However, defect production occurs only upon complete non-adiabatic passage through the QCP. At a 
given time $t$, this holds for $0\ll p_x\ll eEt$, so the number of quenched systems scales $\sim tE$.
Combining these, the Kibble-Zurek mechanism thus also predicts the $tE^{3/2}$ scaling of the total defect 
density for Eq. \eqref{diraceq}, similarly to Eq. \eqref{defectdensity}, linking the non-linear transport in graphene 
 to critical phenomena.
(Quantum critical transport from a different perspective was already studied in Ref. \onlinecite{fritz}.)

Therefore, the low field, perturbative response is dominated by interband contributions, and can be regarded as a manifestation of 
Zitterbewegung. With increasing field, a large number of electron-hole pairs are created, and intraband processes take over, 
producing non-linear transport.
The crossover is determined by the dimensionless time-scale, after comparing our system to the LZ model as\cite{vitanov}
\begin{equation}
\tau_{cross}=\sqrt{\frac{v_FeEt^2}{\hbar}}.
\label{crosstime}
\end{equation}
For $\tau_{cross}\ll 1$, no level crossing occurs, and we can use perturbation theory to estimate the current, therefore we are in 
the Kubo regime. 
The Kibble-Zurek mechanism defines the freeze-out time\cite{zurek,damski} by the instant $\hat t$ when the system leaves the
adiabatic regime and enters into the impulse one, namely $\hat t=\hbar/v_FeE\hat t$, and the Kibble-Zurek form of the defect 
density requires 
complete transit through the QCP, 
$t\gg \hat t$ ($\Leftrightarrow\tau_{cross}\gg 1$).
In the LZ language, level crossing is completed for $\tau_{cross}\gg 1$, the number of pairs created is 
non-perturbative in the electric field, and we can use the probability of LZ tunneling for the current.

So far we have discussed the real time evolution of the current after the switch-on of the electric field, summarized in Table 
\ref{tab1}. In ideal clean graphene, 
for long enough times, Bloch oscillation would set in due to the underlying HCL structure. In reality, the time $t$ 
must be 
replaced, in the spirit of the Drude theory, by an appropriate scattering time\cite{lewkowicz} (due to phonons or impurities), 
$\tau_{sc}$, or in ballistic 
samples, by the 
ballistic flight time from the finite flake size, $\tau_b=L_x/v_F$\cite{allor}. 
The observation of non-linear electric transport requires, from Eq. \eqref{crosstime}, an electric field as 
\begin{equation}
E>E_c=\hbar/v_Fe \tau^2,
\end{equation}
where $\tau=\min(\tau_{sc},\tau_b,\tau_\Delta)$ is the shortest of the additional restricting time scales (with $\tau_\Delta$ 
defined below).
Ballistic transport on the (sub)$\mu$m scale implies $\tau\sim 0.1-1$~ps, giving $E_c\sim 10^3-10^5$~V/m\cite{moser}.
The
measured current is expected to show a change of slope as a function of the electric field in the crossover region, and an
extended electric field window would be required to reveal the non-integer exponent, as shown in Fig. \ref{nlfig}.
It is important to emphasize that in both regions, the current is related to $n_p(t)$, thus even the linear response regime
witnesses pair production.

In the presence of a small mass gap, the above results  need to be modified. The perturbative regime is characterized 
by exponentially activated behaviour due to the gap, and the current is exponentially suppressed at low temperatures ($T\ll\Delta$) 
as $j\sim E\exp(-\Delta/T)$, as in normal semiconductors. On the other hand, for strong electric field, we can still use the analogy 
to LZ tunneling as 
\begin{equation}
\langle j_x\rangle(t)=\frac{2e^2E}{\pi^2\hbar }\sqrt{\frac{v_FeEt^2}{\hbar}}\exp\left(-\frac{\pi\Delta^2}{\hbar v_FeE}\right).
\end{equation}
Non-linear transport sets in for $E>\pi\Delta^2/\hbar v_Fe$, which defines a new timescale for $E_c$ as 
$\tau_\Delta=\hbar/\Delta\sqrt\pi$.

In general, the non-linear current for $d+1$ dimensional ($d=1$, 2, 3) Dirac 
electrons\cite{gavrilov} is
$\langle j_x\rangle(t)\sim tE^{(d+1)/2}\exp(-{\pi\Delta^2}/{\hbar v_FeE})$.
For $d=1$, a good realization would be carbon nanotubes (rolled up graphene sheet), whose "non-linear" response is still linear 
($j\sim E$), only the non-trivial exponential factor with a possible gap reports about non-perturbative effects\cite{andreev}.
The $d=3$ case could be realized among the bulk electrons of Bi, possessing a band-gap $\sim 0.015$~eV.

These results are also relevant for other systems with possible Dirac fermions such as the organic
conductor\cite{katayama} $\alpha$-(BEDT-TTF)$_2$I$_3$ with a tilted Dirac cone.
Dirac fermions can be realized in cold atoms in an appropriate 
optical lattice (half filled HCL, Kagome and triangular lattices),
without any source of dissipation or scattering. The momentum distribution, Eq. \eqref{pairprod} reveals the effect of the driving 
electric field before Bloch oscillations set in\cite{dahan}. 
The pairs created increase the energy of the system as $\sim t^2E^{5/2}$, which, together with the momentum distribution of Bloch 
states, can be measured after releasing the trap. This could be a first direct experimental observation of the Schwinger mechanism as 
well.


\textit{Note added} Recently we became aware of a related work\cite{rosenstein}. Overlapping results are in agreement.

\begin{acknowledgments}
We thank T. Cohen, D. McGady and P. Thalmeier for stimulating discussions and comments, 
and support by the Hungarian Scientific Research Fund No. K72613 and by the Bolyai program of the 
Hungarian Academy of Sciences.
\end{acknowledgments}

\bibliographystyle{apsrev}
\bibliography{refgraph}
\end{document}